\pdfoutput=1
\documentclass[sigconf,natbib=true]{acmart}

\AtBeginDocument{%
  \providecommand\BibTeX{{%
    \normalfont B\kern-0.5em{\scshape i\kern-0.25em b}\kern-0.8em\TeX}}}

\renewcommand\footnotetextcopyrightpermission[1]{}

\usepackage[T1]{fontenc}

\usepackage[hang,flushmargin]{footmisc}

\usepackage{natbib}
\usepackage{hyperref}
\usepackage{url}            % simple URL typesetting
\usepackage{booktabs}       % professional-quality tables
\usepackage{amsfonts}       % blackboard math symbols
\usepackage{nicefrac}       % compact symbols for 1/2, etc.
\usepackage{microtype}      % microtypography
\usepackage{amsmath}
\usepackage{enumitem}
\usepackage{graphicx}       % resizebox
\usepackage{caption}
\usepackage{subcaption}
\usepackage{threeparttable}
\usepackage{color, colortbl}
\usepackage{threeparttable,booktabs}
\usepackage{multirow}
\usepackage{balance}
\usepackage{todonotes}

\newcommand\ignore[1]{}

\usepackage{multirow, makecell}
\usepackage{xcolor}
\usepackage{colortbl}
\usepackage{dsfont}

\usepackage[frozencache=true,cachedir=minted-cache]{minted} 

\definecolor{LightGray}{gray}{0.9}
\PassOptionsToPackage{dvipsnames}{xcolor}
\usepackage{ragged2e}

\newcommand{\gptThreeBold}{\textbf{GPT-3.5}$_\text{\bf Turbo}$}{}
\newcommand{\gptThree}{GPT-3.5$_\text{Turbo}$}{}

\title{LLMs Can Patch Up Missing Relevance Judgments in Evaluation}

\author{Shivani Upadhyay \quad Ehsan Kamalloo \quad Jimmy Lin \\[1ex]
David R. Cheriton School of Computer Science,\\
University of Waterloo, Canada \\[1ex]
\texttt{\{sjupadhyay, ekamalloo, jimmylin\}@uwaterloo.ca}
}

\begin{document}

\renewcommand{\shortauthors}{Upadhyay et al.}
\pagestyle{empty}

\begin{abstract}
%% Actual
Unjudged documents or {\em holes} in information retrieval benchmarks are considered non-relevant in evaluation, yielding no gains in measuring effectiveness.
However, these missing judgments may inadvertently introduce biases into the evaluation as their prevalence for a retrieval model is heavily contingent on the pooling process.
Thus, filling holes becomes crucial in ensuring reliable and accurate evaluation.
Collecting human judgment for all documents is cumbersome and impractical.
In this paper, we aim at leveraging large language models (LLMs) to automatically label unjudged documents.
Our goal is to instruct an LLM using detailed instructions to assign fine-grained relevance judgments to holes.
To this end, we systematically simulate scenarios with varying degrees of holes by randomly dropping relevant documents from the relevance judgment in TREC DL tracks.
Our experiments reveal a strong correlation between our LLM-based method and ground-truth relevance judgments.
Based on our simulation experiments conducted on three TREC DL datasets, in the extreme scenario of retaining only 10\% of judgments, our method achieves a Kendall $\tau$ correlation of 0.87 and 0.92 on an average for Vicuña-7B and {\gptThree} respectively.
\end{abstract}
%\keywords{Evaluation, Relevance Assessment, In-context learning, Large Language Models}
\maketitle

\section{Introduction}
In Information Retrieval (IR), evaluation follows the decades-old Cranfield paradigm \cite{CleverCran} where test collections consist of a set of queries given a text corpus and their relevance judgments.
In the early days, the paradigm assumed that relevance judgments must be complete, i.e. for each query, all documents in the corpus were reviewed.
As the collection size grew over time, the completeness assumption of relevance judgments became increasingly impossible to maintain in the benchmarks.
A feasible strategy to overcome this issue is a process known as {\em pooling}, wherein the top results from one or more retrieval systems are selected for manual judgment.
Pooling heavily rests on the depth of the selected documents, and the choice of retrieval systems.
However, these factors may introduce inadvertent artifacts into which documents are judged and which ones are not, thereby undermining the reliability of the evaluation.
For example, if the retrieval systems leveraged for relevance judgment were all based on BM25 \cite{10.1561/1500000019}, chances are test models that use BM25 score higher in the test collection.
To reduce such risks, we often resort to diversifying retrieval systems for judgment and deeper document selection.
Yet, these solutions come at the cost of more manual labour and slower judgment time, thus making data construction more expensive.

Overall, ground-truth relevance judgments are inevitably {\em incomplete} in IR test collections \cite{10.1145/1008992.1009000}.
Widely used IR metrics such as nDCG@$k$, MAP, or Pr@$k$ either ignore unjudged documents or deem them non-relevant \cite{radlinski2010comparing}.
A major caveat here is to miss relevant but not judged documents \cite{rocketqa} that yield no gain in evaluation.
Also, various retrieval models may have different degrees of sparsity in their judgments, rendering their comparison unfair, particularly for models returning many unjudged documents \cite{arabzadeh2022shallow}.
These shortcomings result in an inaccurate assessment of retrieval models, giving us a misleading sense of progress \cite{armstrong2009improvements,yang2019critically}.

In this paper, we aim to automate the relevance judgment process to minimize human involvement to reduce costs and speed up data collection.
For this purpose, we adopt large language models (LLMs) \cite{brown2020language,opt,black-etal-2022-gpt,llama2} whose strong capabilities to follow instructions with only a few examples \cite{instructgpt,gpt4,bai2022training} show great potential in accomplishing our goal.
In particular, we guide an LLM by providing detailed instructions and carefully crafted examples to assign a relevant label for a given document and query.
Unlike previous studies \cite{10.1145/3539618.3592032} that focused on binary labels, our LLM assessor is instructed to follow TREC guidelines and assign fine-grained relevance labels, which is particularly useful for numerous TREC datasets including TREC DL tracks \cite{craswell2020overview, craswell2021overview,craswell2022overview}.
We simulate varying degrees of unjudged documents ({\em holes}) by randomly removing gold relevance judgments. Then, we run our LLM assessor to patch up the holes.
Our assessor works with open-source LLMs such as Vicuña-7B \cite{zheng2023judging} as well as proprietary ones including {\gptThree}.

We find that LLMs of various sizes are often capable of acting as TREC assessors, showing a strong correlation with gold relevance judgments.
We hope our LLM-based evaluation framework lays the groundwork for a fully automated and robust relevance judgment to eliminate the biases arising from the presence of holes in the data.
Our key contributions can be summarized as follows:

\begin{itemize}[leftmargin=0.5cm]
    \item We introduce an LLM-based framework for filling up the holes in IR test benchmarks.
    \item We extensively evaluate our LLM assessor under various degrees of holes across several TREC DL datasets and find that LLMs exhibit strong correlations with ground-truth relevance judgments.
    \item Our framework will be shipped as an easy-to-use tool for researchers and practitioners to measure the effectiveness of retrieval systems with no holes in relevance judgments.
\end{itemize}

\section{Related Work}

\paragraph{Metrics for Incomplete Judgments:}
Metrics such as nDCG@$k$, MAP, and Pr@$k$ require the availability of relevance for the top k documents \cite{radlinski2010comparing}. Upon increase in the prevalence of usage for such metrics, accurate handling of the unjudged documents turned out to be a vital factor in the assessment of retrieval systems. Simple approaches such as removal of the unjudged documents or treating them as non-relevant gained much popularity however, they both still miss the true essence of the complete judgment. \cite{yilmaz2008simple} presents the study of sensitivity to erroneous judgments for nDCG metrics. Statistical methods based on the sampling and prior are presented in \cite{frobe2023bootstrapped, frobe2022noise} for filling up the holes for metric calculations.

\paragraph{Automated Relevance Judgment:}
The high effort needed for manual test preparation for large-scale testing has always encouraged researchers to move towards an automated relevance system that could provide us with a comparable correlation with the manual test assessments \cite{dietz2022wikimarks, dietz2020humans}. Usage of LLMs for complete or partial (filling only holes) relevance judgments has been tested under various scenarios such as zero-shot prompting, one-shot learning, describing a role, varying description/aspects, and the number of judges, and LLMs tend to provide comparable correlation results with actual judgment/metric in most of the settings \cite{faggioli2023perspectives,thomas2023large, 10.1145/3539618.3592032}.

\paragraph{LLMs for Text Annotation:} 
There has been continuous research for leveraging LLMs and substituting manual text annotations. Proprietary models such as ChatGPT and GPT-4 have demonstrated a great understanding of textual data and their potential for substituting manual labellers \cite{zhu2023can,savelka2023can}.
Even, the open-source LLMs have provided results comparable with proprietary models in the task of text annotations \cite{alizadeh2023open}.

\section{Methodology}
Our main goal in this paper is to harness the powerful capabilities of LLMs to fill up the missing judgments present in ground-truth relevance judgments, thereby mitigating the impact of holes in IR test benchmarks.

%\subsection{TREC DL track evaluations}

\begin{figure}[t]
\tiny
\justifying
\begin{minted}[fontsize=\footnotesize, frame=lines, frame=single,linenos=false,breaklines,breaksymbol=,escapeinside=||,bgcolor=LightGray]{text}
You are an expert judge of content. Using your internal knowledge and simple commonsense reasoning, try to verify if the passage is relevant to the query.
Here, "0" represents that the passage has nothing to do with the query, "1" represents that the passage seems related to the query but does not answer it, "2" represents that the passage has some answer for the query, but the answer may be a bit unclear, or hidden amongst extraneous information and "3" represents that the passage is dedicated to the query and contains the exact answer.

{examples}

Provide an explanation for the relevance and give your answer from one of the categories 0, 1, 2 or 3 only. One of the categorical values is compulsory in the answer.

Instructions: Think about the question. After explaining your reasoning, provide your answer in terms of 0, 1, 2 or 3 categories. Only provide the relevance category on the last line. Do not provide any further details on the last line.

###

Query: {query}
Passage: {passage}

Explanation:
\end{minted}

\caption{Instructions provided in the prompts (without in-context learning).}
\vspace{-0.25cm}
\label{fig:prompt-struture}
\end{figure}

% The key assumption under the Cranfield \cite{CleverCran} testing paradigm is that the ground-truth judgments are complete. 
The Cranfield-based test collections when used for evaluation in the TREC Deep Learning (DL) track \cite{craswell2020overview, craswell2021overview,craswell2022overview} are deemed comprehensive and assumed that every related document for the query has been assessed. However, this assumption cannot be held in practice and thus, we select a subset of documents using pooling for human judgment that leads to the presence of unjudged documents. Please refer to Table \ref{tab:dl-tracks} which provides statistics about TREC DL 2019, 2020 and 2021 used for the evaluations. Mainly, we focused passage ranking task for this paper.

In computing evaluation metrics like nDCG@$k$, it's a common practice to term unjudged documents as non-relevant. As a result, retrieval models that select documents not included in the pooling process do not gain any positive benefit for their predictions.

\begin{figure}[t]
\tiny
\justifying
\begin{minted}[fontsize=\footnotesize, frame=lines, frame=single,linenos=false,breaklines,breaksymbol=,escapeinside=||,bgcolor=LightGray]{text}
Following are some of the examples of relevance categorizations for different categories:

...
###

Query: when did rock n roll begin?
Passage: With a name based on a Muddy Waters song The Rolling Stones formed in 1962. ...
Relevance category: 0

###

Query: who is rep scalise?
Passage: Senators: Scalise Made 'Grave Mistake' Speaking to White Supremacist Group. Two senators ...
Relevance category: 1

###

Query: when did rock n roll begin?
Passage: Who Really Invented Rock 'n' Roll. This is a fair and clever summary of what ...
Relevance category: 2

###

Query: what amino produces carnitine
Passage: Everyone needs lysine, though some people benefit from it more than others. Lysine ...
Relevance category: 3

...

\end{minted}

\caption{Examples provided in the prompt for in-context learning.}
\vspace{-0.25cm}
\label{fig:prompt-examples}
\end{figure}

% Now to fill these (newly formed) holes we use LLMs and understand how accurately they can re-fill them. 
\renewcommand\theadfont{\normalsize}

\begin{table*}[t]
\caption{Summary statistics for TREC DL 2019, 2020 and 2021 tracks.}
  \begin{tabular}{c|c|c|c|c|c}
    \toprule
    \multirow{2}{*}{\textbf{DL track}} & \multicolumn{5}{c}{\footnotesize\textbf{Total counts for}}\\
    &\textbf{submissions}&\textbf{topics}&\textbf{relevance label 1}&\textbf{relevance label 2}&\textbf{relevance label 3}\\
    \midrule
    TREC 2019 & 36 & 200 & 1601 & 1804 & 697\\
    TREC 2020 & 59 & 200 & 1940 & 1020 & 646\\
    TREC 2021 & 62 & 477 & 3063 & 2341 & 1086\\
  \bottomrule
  \end{tabular}
  \label{tab:dl-tracks}
    \vspace{-0.1cm}
\end{table*}

\begin{figure*}[t]
%\vspace{-0.5cm}
  \centering
  \begin{subfigure}{0.33\textwidth}
    \centering
    \includegraphics[scale=0.32]{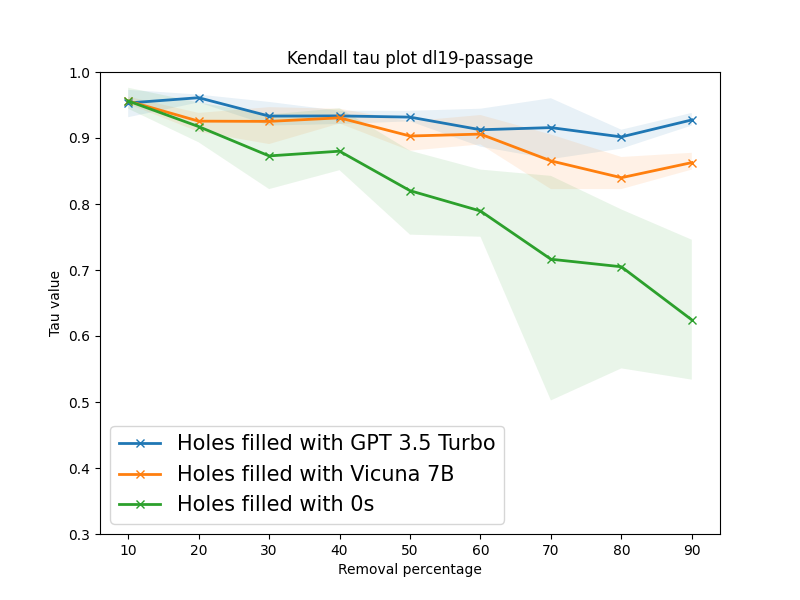}
    \vspace{-0.1cm}
    \caption{Qrel: TREC 2019}
    \label{fig:image1}
  \end{subfigure}
  \hfill
  \centering
  \begin{subfigure}{0.33\textwidth}
    \centering
    \includegraphics[scale=0.32]{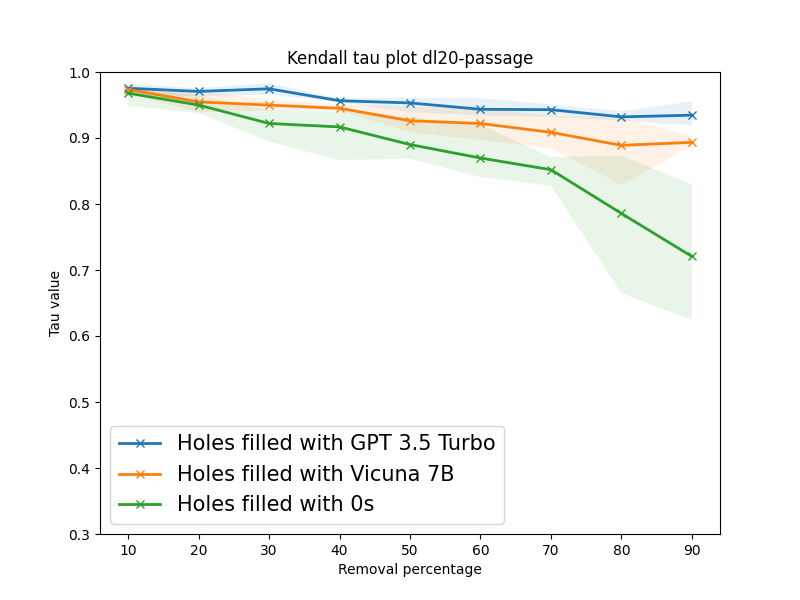}
    \vspace{-0.1cm}
    \caption{Qrel: TREC 2020}
    \label{fig:image2}
  \end{subfigure}
  \hfill
  \centering
  \begin{subfigure}{0.33\textwidth}
    \centering
    \includegraphics[scale=0.32]{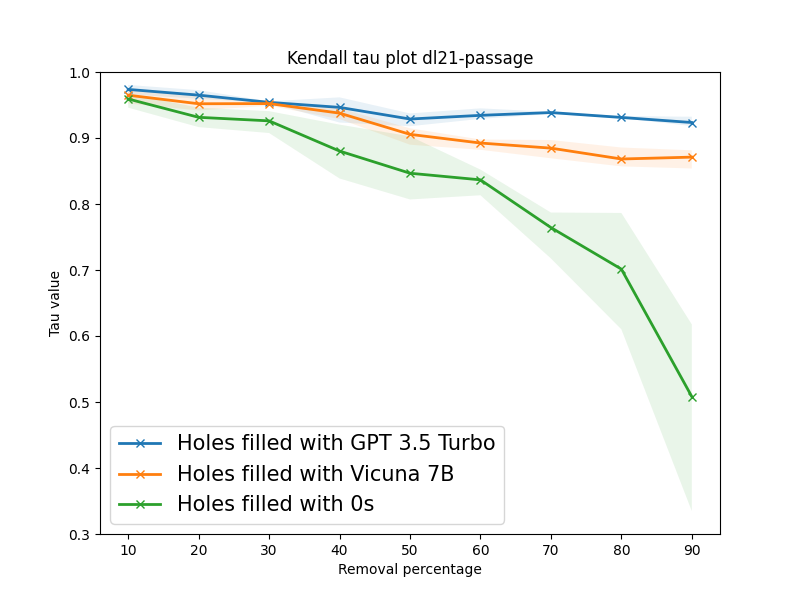}
    \vspace{-0.1cm}
    \caption{Qrel: TREC 2021}
    \label{fig:image3}
  \end{subfigure}
   \vspace{-0.5cm}
  \caption{Line plots for comparing Kendall $\tau$ value when the patched incomplete judgment file compared with the complete judgment (ground truth). Three scenarios are being compared for each TREC DL track:\ holes plugged with {\gptThree} (marked with Blue), Vicuña-7B (marked with Orange), and non-relevant label (marked with Green).}
  \label{fig:combined}
  \vspace{-0.2cm}
\end{figure*}

%\subsection{Filling up the holes}

To overcome this shortcoming, we propose an evaluation framework where the holes are explicitly filled before computing the metric. Thus, filling up holes with correct relevant labels strengthens the Cranfield assumption and leads to more accurate and reliable assessments.

To fill up the holes we utilize LLMs to understand the relation between query and passage, and then use the provided label for filling the holes. 
% Firstly, we pre-process the incomplete judgment (i.e. one with unjudged documents) for fetching the relevant labels. 
Here, LLMs are used to understand the relevance level for the given query-passage pair and generate the label accordingly. For this step, we guide the LLMs with few-shot prompting \cite{brown2020language}, which uses two examples of each relevance label from the judged query-passage pair (as illustrated in Figure~\ref{fig:prompt-examples}). Our framework assumes that we will have at least two pairs of assessed query documents available to make the model understand the task. This assumption is fairly reasonable and easy to attain in practice. Figure~\ref{fig:prompt-struture} presents the complete prompt. Once we have generated relevant labels for incomplete judgments, we will use them to fill the holes.

Equation~\eqref{eq:Q} illustrates our framework's working. Here $\text{Q}$ denotes the incomplete judgment that needs to be filled, $\text{Q}^{'}$ represents the modified judgment, $q^{a}$ refers to assessed pairs, $q^{h}$ represents holes, $sample(q^{a})$ denotes the two sample instances from each label for passing it as examples with the prompt and the output of the function $eval\left(q^{h}, sample(q^{a})\right) \epsilon \left[0, 1, 2, 3\right]$ will the relevance label generated using LLM.

\begin{equation}\label{eq:Q}
  Q'(q) =
  \begin{cases}
    Q(q) & q \text{ } \epsilon \text{ } q^{a}, \\
    eval\left(q^{h}, sample(q^{a})\right) & q \text{ } \epsilon \text{ } q^{h}, \\
  \end{cases}
\end{equation}

Thus, the newly generated relevance label for holes depends on the query-passage pair and the sampled examples passed for in-context learning for the LLM.

\section{Experimental Setup}
% In this section, we examine the effectiveness of our proposed LLM-based evaluation framework for filling holes in relevance judgments.

\paragraph{Data:}
In our experiments, we used TREC DL \textit{2019}, \textit{2020} and \textit{2021} passage ranking qrels for preparing synthetic judgments that contain holes.\footnote{We excluded TREC DL 2022 because their higher number of judgments exceeds our allocated budget limit.} A TREC DL judgment includes \textit{qid}: unique id given to the MS MARCO query, \textit{pid}: unique id given to the passage, \textit{rating}: relevance label assigned to the passage. Roughly each of the TREC qrels includes query-passage pairs belonging to all four relevance classes (i.e. 3: perfectly relevant, 2: highly relevant, 1: related, and 0: irrelevant).

%\subsection{Experimental Setup}

\paragraph{Preparing Synthetic Incomplete Judgments:}
To test the effectiveness of our framework, we synthetically prepare incomplete judgments. We sample certain percentages of the complete judgments from each relevant (i.e. label: 1, 2, and 3) label and they are marked as holes for our synthetic incomplete judgments. The sampling percentages that we used for formulations are \{10, 20, 30, 40, 50, 60, 70, 80, 90\}. 
For instance, let the complete judgment include $a$ query-passage pairs of label 3, $b$ query-passage pairs of label 2, $c$ query-passage pairs of label 1, and $d$ query-passage pairs of label 0, then for the formulation of synthetic judgment, we will be removing $X$\% of judgments from $a, b \text{ and } c$.

\paragraph{LLMs Used for Evaluations:}
For evaluations, we have results using the OpenAI proprietary model {\gptThree}\footnote{\url{https://platform.openai.com/docs/models/gpt-3-5-turbo}} and the open-source model Vicuña-7.\footnote{\url{https://huggingface.co/lmsys/vicuna-7b-v1.5}} Apart from runs that required API calls for \gptThree, we have utilized a single NVIDIA RTX A6000 GPU for all Vicuña-7 inference runs unless stated otherwise.
% \todo{GPU utilization info}

\section{Results and Discussion}

\paragraph{Correlation with Complete Judgment:}
Figure \ref{fig:combined} represents a line plot comparison for Kendall $\tau$ correlation values in ranking retrieval systems based on metric nDCG@10 with the rankings obtained using original ground truth judgments. For synthetically generated incomplete judgments, three trials were performed and the figure displays line plots with mean $\tau$ value aggregated over trials. The coloured area surrounding the line presents the variance among the trials for each experiment. Vicuña-7B and {\gptThree} present a strong correlation\footnote{rule-of-thumb: correlation value that is greater than 0.8 can be considered as a ``strong'' as per statistics literature.} with the original rankings, however, we see gradual degradation in the correlation when the holes are simply termed as non-relevant (0 labels). 

Table \ref{tab:tau} provides the mean Kendall $\tau$ correlation values for the extreme scenarios where only 10\% of the ground truth assessed judgments were retained. Notably, Vicuña and {\gptThree} provide us with an average 25\% and 30\% higher correlation, respectively when compared with the strategy where we simply fill holes with the non-relevant (0) label.

\begin{figure}[t]
%    \vspace{-0.3cm}
  \centering
   \includegraphics[scale=0.35]{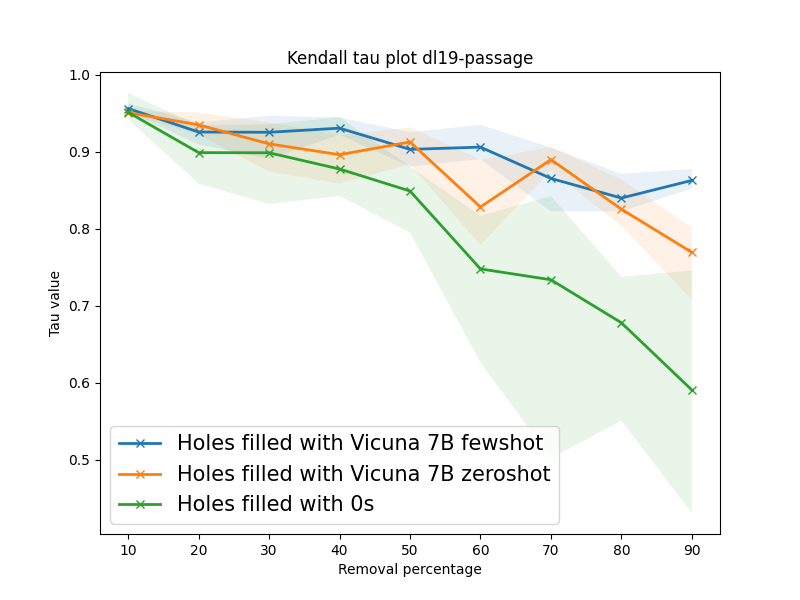}
   \vspace{-0.3cm}
   \caption{Line plots for comparing few-shot (marked with Blue) vs. zero-shot (marked with Orange) prompting for filling up the holes in TREC DL 2019.}
   \label{fig:zero_vs_few}
%\vspace{-0.2cm}
\end{figure}

\begin{table}[t]
\caption{Kendall $\tau$ comparisons between nDCG@10 scores assessed using original judgments and judgments with 90\% holes when evaluated with our framework. Table \ref{tab:dropped-labels} provides further statistics about the synthetically curated incomplete judgment.}
\vspace{-0.3cm}
  \begin{tabular}{c|c|c|c}
    \toprule
    \multirow{2}{*}{\textbf{DL track}} & \multicolumn{3}{c}{\footnotesize\textbf{Kendall $\mathbf{\tau}$ when holes plugged with}}\\
    &{\gptThreeBold}&\textbf{Vicuña-7B}&\textbf{Non-relevants}\\
    \midrule
    TREC 2019 & 0.927 & 0.862 & 0.624\\
    TREC 2020 & 0.934 & 0.893 & 0.720\\
    TREC 2021 & 0.923 & 0.871 & 0.508\\
  \bottomrule
  \end{tabular}
  \label{tab:tau}
% \vspace{-0.5cm}
\end{table}

\paragraph{Zero-Shot vs Few-Shot Prompting:}
Figure \ref{fig:zero_vs_few} presents the comparison between zero-shot and few-shot prompting for filling the holes. Here, we are comparing the TREC DL 2019 correlation results of ground truth judgments with the judgments filled using different techniques. In each experiment, we performed three trials and the bold line presents the value of mean $\tau$. The shaded area shows the variance among different trials. The zero-shot prompting demonstrates comparable effectiveness with few-shot, however with zero-shot, we observe huge variance compared to few-shot. Especially $\tau$ for some of the experiments goes even lower than 0.8 and thus deteriorates the correlation.

\begin{table}[t]
\caption{Summary of the dropped relevance labels for synthetically created incomplete judgment with 90\% of holes.}
\vspace{-0.1cm}
  \begin{tabular}{c| r  r  r }
    \toprule
    \multirow{2}{*}{\textbf{DL track}} & \multicolumn{3}{c}{\footnotesize\textbf{Count for dropped relevances}}\\
    &\textbf{label 1}&\textbf{label 2}&\textbf{label 3}\\
    \midrule
    TREC 2019 & 1440 & 1623 & 627\\
    TREC 2020 & 1746 & 918 & 581\\
    TREC 2021 & 2756 & 2106 & 977\\
  \bottomrule
  \end{tabular}
  \label{tab:dropped-labels}
% \vspace{-0.1cm}
\end{table}

\paragraph{Data Contamination Testing:}
A major concern emerging with the LLMs usage is data leakage which might lead to LLMs cheating and generating higher results.
Given the public nature of TREC DL datasets, similar concerns are bound to be raised for the mentioned framework results.
Thus, to verify if these metrics suffered from LLM cheating or not, we performed a small experiment with DL TREC 2023 qrels as it has the highest chance of not being seen at training time.
As showcased in Table \ref{tab:bm25-comp}, we were able to achieve effectiveness for the BM25 retriever system with incomplete qrels using our framework that is comparable with ground-truth qrels metrics.
Here, the incomplete qrels represent the synthetically generated incomplete qrels with 90\% unjudged documents.
Hence, we can interpret that the LLMs have the capabilities of understanding the query-passage pairs and assigning them with appropriate relevance labels and not just result in data leakage.

\begin{table}[t]
\caption{Comparison for BM25 results evaluated using ground-truth qrels and the incomplete qrels filled using \gptThree and Vicuña-7B. Here, \textit{GT} represents evaluation using the ground-truth qrels and \textit{IC} represents evaluation with the incomplete qrels.}
\vspace{-0.2cm}
  \begin{tabular}{p{3.5cm}|c c}
    \toprule
    \textbf{Experiment} & \textbf{nDCG@10} & \textbf{MAP}\\
    \midrule
    BM25 (GT) & 0.2627 & 0.0793\\
    BM25 (IC) + \gptThree & 0.2721 & 0.0861\\
    BM25 (IC) + Vicuña-7B & 0.1553 & 0.0623\\
  \bottomrule
  \end{tabular}
  \label{tab:bm25-comp}
\vspace{-0.25cm}
\end{table}

\section{Conclusion}
Incomplete ground-truth relevance judgments will create holes in evaluating IR models.
The presence of these holes may inadvertently bias the evaluation results, thus leading to misleading conclusions.
In this work, we present a simple yet effective evaluation framework to patch up the holes by instructing LLMs to predict fine-grained relevance labels for holes.
We find that a 7B open-source LLM as well as the full-fledged proprietary {\gptThree} both exhibit strong correlations with human-labelled relevance judgments.
Our framework will be made available to the community to offer necessary tools to patch up holes in testing retrieval models. 

We did not address {\em imperfect} relevance judgments (existing relevance labels that are not correct) in this work.
Imperfect judgments can arise due to many reasons such as noise or erroneous judgments. How to automatically amend these judgments is a challenge that future work can focus on.
Another interesting direction for future work is to cast existing relevance judgments as preference data and use them to fine-tune LLMs using Reinforcement Learning from Human Feedback (RLHF) to build a specialized LLM assessor.

\section*{Acknowledgements}

This research was supported in part by the Canada First Research Excellence Fund and the Natural Sciences and Engineering Research Council (NSERC) of Canada.
We'd also like to thank Microsoft for providing access to OpenAI LLMs via Azure.

\bibliographystyle{ACM-Reference-Format}
\bibliography{custom}

\appendix









\end{document}